%
\let\includefigures=\iftrue
%
%
%
%
%
\input harvmac
\input rotate
\input epsf
\input xyv2
\noblackbox
\includefigures
\message{If you do not have epsf.tex (to include figures),}
\message{change the option at the top of the tex file.}
\def\figin{\epsfcheck\figin}\def\figins{\epsfcheck\figins}
\def\epsfcheck{\ifx\epsfbox\UnDeFiNeD
\message{(NO epsf.tex, FIGURES WILL BE IGNORED)}
\gdef\figin##1{\vskip2in}\gdef\figins##1{\hskip.5in}
\else\message{(FIGURES WILL BE INCLUDED)}%
\gdef\figin##1{##1}\gdef\figins##1{##1}\fi}
\def\DefWarn#1{}

\def\figinsert{\goodbreak\midinsert}
\def\ifig#1#2#3{\DefWarn#1\xdef#1{fig.~\the\figno}
\writedef{#1\leftbracket fig.\noexpand~\the\figno}%
\figinsert\figin{\centerline{#3}}\medskip\centerline{\vbox{\baselineskip12pt
\advance\hsize by -1truein\noindent\footnotefont{\bf
Fig.~\the\figno:} #2}}
\bigskip\endinsert\global\advance\figno by1}
\else
\def\ifig#1#2#3{\xdef#1{fig.~\the\figno}
\writedef{#1\leftbracket fig.\noexpand~\the\figno}%
\global\advance\figno by1} \fi
\def\Title#1#2{\rightline{#1}\ifx\answ\bigans\nopagenumbers\pageno0
\else\pageno1\vskip.5in\fi \centerline{\titlefont #2}\vskip .3in}
\font\caps=cmcsc10

\def\yboxit#1#2{\vbox{\hrule height #1 \hbox{\vrule width #1
\vbox{#2}\vrule width #1 }\hrule height #1 }}
\def\fillbox#1{\hbox to #1{\vbox to #1{\vfil}\hfil}}
\def\ybox{{\lower 1.3pt \yboxit{0.4pt}{\fillbox{8pt}}\hskip-0.2pt}}

\def\rightarrowbox#1#2{
  \setbox1=\hbox{\kern#1{${ #2}$}\kern#1}
  \,\vbox{\offinterlineskip\hbox to\wd1{\hfil\copy1\hfil}
    \kern 3pt\hbox to\wd1{\rightarrowfill}}}

\def\CO{{\cal O}}

\def\tilde{\widetilde}

           \def\CO{{\cal O}}


\def\dj{\hbox{d\kern-0.347em \vrule width 0.3em height 1.252ex depth
-1.21ex \kern 0.051em}}

\def\Dirac{\,\raise.15ex\hbox{/}\mkern-13.5mu D}
\def\dirac{\,\raise.15ex\hbox{/}\kern-.57em \partial}
\def\aslash{\,\raise.15ex\hbox{/}\mkern-13.5mu A}

\def\shalf{{\ifinner {\textstyle {1 \over 2}}\else {1 \over 2} \fi}}
\def\sshalf{{\ifinner {\scriptstyle {1 \over 2}}\else {1 \over 2} \fi}}
\def\sfourth{{\ifinner {\textstyle {1 \over 4}}\else {1 \over 4} \fi}}
\def\sthreehalfs{{\ifinner {\textstyle {3 \over 2}}\else {3 \over 2} \fi}}
\def\sdhalfs{{\ifinner {\textstyle {d \over 2}}\else {d \over 2} \fi}}
\def\sdmtwohalfs{{\ifinner {\textstyle {d-2 \over 2}}\else {d-2 \over 2} \fi}}
\def\sdmasonehalfs{{\ifinner {\textstyle {d+1 \over 2}}\else {d+1 \over 2} \fi}}
\def\sdmasthreehalfs{{\ifinner {\textstyle {d+3 \over 2}}\else {d+3 \over 2} \fi}}
\def\sdmastwohalfs{{\ifinner {\textstyle {d+2 \over 2}}\else {d+2 \over 2} \fi}}

\def\ncell{n_{\rm cell}}
\def\scell{S_{\rm cell}}


 \lref\adscft{
  J.~M.~Maldacena,
  ``The large N limit of superconformal field theories and supergravity,''
  Adv.\ Theor.\ Math.\ Phys.\  {\bf 2}, 231 (1998)
  [Int.\ J.\ Theor.\ Phys.\  {\bf 38}, 1113 (1999)]
  [arXiv:hep-th/9711200].
 S.~S.~Gubser, I.~R.~Klebanov and A.~M.~Polyakov,
  ``Gauge theory correlators from non-critical string theory,''
  Phys.\ Lett.\  B {\bf 428}, 105 (1998)
  [arXiv:hep-th/9802109].
 E.~Witten,
  ``Anti-de Sitter space and holography,''
  Adv.\ Theor.\ Math.\ Phys.\  {\bf 2}, 253 (1998)
  [arXiv:hep-th/9802150].
  }
  
  \lref\Matrix{
  T.~Banks, W.~Fischler, S.~H.~Shenker and L.~Susskind,
  ``M theory as a matrix model: A conjecture,''
  Phys.\ Rev.\  D {\bf 55}, 5112 (1997)
  [arXiv:hep-th/9610043].
}

\lref\Hayden{
  P.~Hayden and J.~Preskill,
  ``Black holes as mirrors: quantum information in random subsystems,''
  JHEP {\bf 0709}, 120 (2007)
  [arXiv:0708.4025 [hep-th]].
}

\lref\Sekino{
  Y.~Sekino and L.~Susskind,
  ``Fast Scramblers,''
  JHEP {\bf 0810}, 065 (2008)
  [arXiv:0808.2096 [hep-th]].
}

\lref\Susskindnew{
  L.~Susskind,
  ``Addendum to Fast Scramblers,''
  arXiv:1101.6048 [hep-th].
}

\lref\Damour{
  T.~Damour,
  ``Black Hole Eddy Currents,''
  Phys.\ Rev.\  D {\bf 18}, 3598 (1978).
}

\lref\Thorne{
  K.~S.~.~Thorne, R.~H.~.~Price and D.~A.~.~Macdonald,
  ``Black Holes: The Membrane Paradigm,''
{\it  New Haven, USA: Yale Univ. Pr. (1986) 367p}
}

\lref\Susskindbook{
  L.~Susskind and J.~Lindesay,
  ``An introduction to black holes, information and the string theory
  revolution: The holographic universe,''
{\it  Hackensack, USA: World Scientific (2005) 183 p}
}

  \lref\itzhaki{
  N.~Itzhaki, J.~M.~Maldacena, J.~Sonnenschein and S.~Yankielowicz,
  ``Supergravity and the large N limit of theories with sixteen
  supercharges,''
  Phys.\ Rev.\  D {\bf 58}, 046004 (1998)
  [arXiv:hep-th/9802042].
}

\lref\gl{
 R.~Gregory and R.~Laflamme,
  ``Black strings and p-branes are unstable,''
  Phys.\ Rev.\ Lett.\  {\bf 70}, 2837 (1993)
  [arXiv:hep-th/9301052].
  R.~Gregory and R.~Laflamme,
  ``The Instability of charged black strings and p-branes,''
  Nucl.\ Phys.\  B {\bf 428}, 399 (1994)
  		  [arXiv:hep-th/9404071].}
		  
\lref\kogan{
 J.~L.~F.~Barbon, I.~I.~Kogan and E.~Rabinovici,
  ``On stringy thresholds in SYM/AdS thermodynamics,''
  Nucl.\ Phys.\  B {\bf 544}, 104 (1999)
  [arXiv:hep-th/9809033].		  }

\lref\optical{
  G.~W.~Gibbons and M.~J.~Perry,
  ``Black Holes And Thermal Green's Functions,''
  Proc.\ Roy.\ Soc.\ Lond.\  A {\bf 358}, 467 (1978).
  G.~Kennedy, R.~Critchley and J.~S.~Dowker,
  ``Finite Temperature Field Theory with Boundaries: Stress Tensor and Surface
  Action Renormalization,''
  Annals Phys.\  {\bf 125}, 346 (1980).
}

\lref\solo{
I.~Sachs and S.~N.~Solodukhin,
  ``Horizon holography,''
  Phys.\ Rev.\  D {\bf 64}, 124023 (2001)
  [arXiv:hep-th/0107173].
  }
  
  \lref\solor{
  S.~N.~Solodukhin,
  ``Entanglement entropy of black holes,''
  arXiv:1104.3712 [hep-th].
	  }
	  
\lref\gib{
G.~W.~Gibbons and C.~M.~Warnick,
  ``Universal properties of the near-horizon optical geometry,''
  Phys.\ Rev.\  D {\bf 79}, 064031 (2009)
  [arXiv:0809.1571 [gr-qc]].
  }	  

\lref\exten{
J.~L.~F.~Barbon and E.~Rabinovici,
  ``Extensivity versus holography in anti-de Sitter spaces,''
  Nucl.\ Phys.\  B {\bf 545}, 371 (1999)
  [arXiv:hep-th/9805143].
}

\lref\ncop{
J.~L.~F.~Barbon and E.~Rabinovici,
  ``On 1/N corrections to the entropy of noncommutative Yang-Mills  theories,''
  JHEP {\bf 9912}, 017 (1999)
  [arXiv:hep-th/9910019].
}

  \lref\beren{
  C.~Asplund, D.~Berenstein and D.~Trancanelli,
  ``Evidence for fast thermalization in the BMN matrix model,''
  arXiv:1104.5469 [hep-th].
  }

  \lref\festuccia{
  G.~Festuccia and H.~Liu,
  ``The arrow of time, black holes, and quantum mixing of large N Yang-Mills
  theories,''
  JHEP {\bf 0712}, 027 (2007)
  [arXiv:hep-th/0611098].
  N.~Iizuka and J.~Polchinski,
  ``A Matrix Model for Black Hole Thermalization,''
  JHEP {\bf 0810}, 028 (2008)
  [arXiv:0801.3657 [hep-th]].
  N.~Iizuka, T.~Okuda and J.~Polchinski,
  ``Matrix Models for the Black Hole Information Paradox,''
  JHEP {\bf 1002}, 073 (2010)
  [arXiv:0808.0530 [hep-th]].
  }

\lref\opme{
J.~L.~F.~Barbon,
  ``Horizon divergences of fields and strings in black hole backgrounds,''
Phys.\ Rev.\  {\bf D50}, 2712-2718 (1994).
[hep-th/9402004].
}

\lref\uschaos{
J.~L.~F.~Barbon and J.~M.~Magan,
  ``Chaotic Fast Scrambling At Black Holes,''
  arXiv:1105.2581 [hep-th].
}

\lref\lstus{
J.~L.~F.~Barbon, C.~A.~Fuertes and E.~Rabinovici,
  ``Deconstructing the little Hagedorn holography,''
  JHEP {\bf 0709}, 055 (2007)
  [arXiv:0707.1158 [hep-th]].
  }



\line{\hfill IFT UAM/CSIC-11-43}

\vskip 0.7cm

\Title{\vbox{\baselineskip 12pt\hbox{}
 }}
{\vbox {\centerline{ Fast Scramblers Of Small Size
 }
\vskip10pt
}}
\vskip 0.5cm

\centerline{$\quad$ {
{\caps Jos\'e L.F. Barb\'on}
 {\caps and}
{\caps Javier M. Mag\'an}
}}
\vskip0.5cm

\centerline{{\sl   Instituto de F\'{\i}sica Te\'orica IFT UAM/CSIC }}
\centerline{{\sl  Campus de Cantoblanco 28049. Madrid, Spain }}
\centerline{{\tt jose.barbon@uam.es}, {\tt javier.martinez@uam.es}}

\vskip1.2cm

\centerline{\bf ABSTRACT}

 \vskip 0.3cm

 \noindent

We investigate various geometrical  aspects of the notion of `optical depth' in the thermal atmosphere of black hole horizons. Optical depth has been proposed as a measure of fast-crambling times in such black hole systems, and the associated optical metric suggests that classical chaos plays a leading role in the actual scrambling mechanism.  We study the behavior of the optical depth with the size of the system and find that AdS/CFT phase transitions with topology change occur naturally as the scrambler becomes smaller than its thermal length. In the context of detailed AdS/CFT models based on D-branes, T-duality implies that small scramblers are described in terms of matrix quantum mechanics.

\vskip 0.2cm

\Date{June  2011}

\vfill





\baselineskip=15pt

\newsec{Introduction}

\noindent

\noindent

Fast scramblers were introduced in 
 \refs{\Hayden, \Sekino}  as systems that saturate causality bounds on the retrieval of information from black holes. It was actually conjectured in \refs{\Sekino} that black holes are the fastest scramblers in nature, and that large-$N$ matrix models share this property as effective microscopic descriptions of their horizon degrees of freedom (see also \refs\Susskindnew). 
 
 The scrambling time scale for black holes is conjectured to be
 \eqn\fsts{
 \tau_s = \beta\,\log(S)\;,
 }
 where $\beta=T^{-1}$ is the inverse Hawking temperature and $S$ is the entropy. This formula should apply both to Schwarzschild black holes, whose size is controlled by $\beta$, and to near-extremal Reissner--Nordstrom black holes, whose size is much smaller than $\beta$.   On the other hand, since large $N$ CFTs at strong coupling have thermal states described by black hole backgrounds in AdS spacetimes, it is tempting to assume that \fsts\ also applies in general to finite-temperature states in such CFTs. 
 
 The fast-scrambling  time scale is clearly faster than the diffusion time scale for a local system with $\CO(1)$ degrees of freedom per thermal length, $\beta$, and size $L$, given by
 \eqn\diffe{
 \tau_{\rm diff} \sim L^2 \,T\sim \beta\,(S)^{2/d}\;,
 }
 where $d$ is the number of spatial dimensions, $S\sim (LT)^d$ is the extensive entropy, and again we assume $\beta = T^{-1}$ to be the only relevant energy scale of the system. In fact, \fsts\ is even faster than the time required to traverse the system at the speed of light:
 \eqn\causa{
 \tau_{\rm causal} \sim L = \beta \,LT\sim \beta \,(S)^{1/d}\;,
 }
 so that a literal application of \fsts\ to any type of horizon cannot be expected to hold. 
 
 A natural compromise for conformal systems, suggested in \refs\Sekino, would have \fsts\ applying at the level of the {\it thermal cell}, i.e. the volume associated to a single thermal length of the system, $V_{\rm cell} = \beta^d$. In this sense, \fsts\ only acquires a non-trivial character when extrapolated to systems with a large number of degrees of freedom, or entropy per thermal cell $S_{\rm cell} \equiv N_{\rm eff} \gg 1$.  
 
 It was recently observed in \refs\uschaos\ that precisely such a time scale appears in complete generality as a kinematical factor in any near-horizon Rindler region, as a causality bound for information to return from the stretched horizon. In other words, the strict reflection time of light signals across the Rindler region of any holographic background takes the form, once translated to CFT variables, and neglecting $\CO(1)$ multiplicative and additive factors,
 \eqn\dealy{
 \tau_{\rm delay}\sim  \tau_*  \sim \beta\,\log(N_{\rm eff})\;.
 }
It was also pointed out in \refs\uschaos\ that \dealy\  can be given a geometrical interpretation as the {\it optical depth} of the thermal atmosphere, i.e. the size of the Rindler region above the stretched horizon, measured in the {\it optical metric} $ds^2_{\rm op} \equiv ds^2 /|g_{00}|$. The optical metric has a universal behavior in the near-horizon region, being  well-approximated by a constant-curvature hyperboloid (see \refs{\optical, \opme, \solo, \solor, \gib} for previous applications of the optical metric). Under these conditions, it was argued in \refs\uschaos\ that a purely kinetic model of scrambling for local point-like probes achieves delocalization over one thermal cell precisely on a time scale of the order of \dealy. The reason is apparent from the hyperbolic nature of the optical metric and the $\CO(1)$ interactions expected at the stretched horizon, producing a scenario quite similar to a hyperbolic billiard, a classic example of hard chaos in classical mechanics. Since the time scale for a free ballistic glide at the speed of light is of order $\tau
 _*$ in the Rindler geometry, the kinetic scrambling is achieved in a  Lyapunov time of a few collisions and hence we can expect $\tau_s \sim \tau_* $ in order of magnitude, provided we consider a single thermal cell. 

On distance scales larger than a thermal cell, no free ballistic paths can occur  in a single step within the hyperbolic geometry, due to the finite optical depth of the Rindler region, and the reflecting nature of the asymptotic metric in a holographic background (absence of true Hawking radiation). This implies that the kinetic scrambling behaves like a standard diffusive process over larger regions, with basic time step $\tau_{\rm delay}$, leading to a final scrambling time of order 
\eqn\fins{
\tau_s \sim (n_{\rm cell})^{2/d} \, \beta \,\log(N_{\rm eff})\;,
}
where $n_{\rm cell} \equiv VT^d$ is the number of thermal cells in the system.   This result follows from a concrete kinetic model with minimal assumptions, and is compatible with causality constraints \causa, as well as with
the general expectation that, on very large length scales, any CFT thermal state should scramble diffusively, as corresponds to a local system. From this point of view, the anomalous behavior of a single thermal cell \fins\ comes to be viewed as a large-$N$ effect of the CFT, a peculiar {\it retardation} effect in the complete scrambling of $\CO(N_{\rm eff})$ degrees of freedom, as compared to the local thermalization time of $\CO(\beta)$ which, measured in terms of local operators, only affects $\CO(1)$ degrees of freedom.

In this note we look more carefully at the effects of finite horizon size in these results. We begin in the next section with a review of the estimate of \refs\uschaos\ for the optical depth of a generic holographic thermal state in the absence of finite-size effects.  In section 3 we consider finite-size effects in the bulk geometry, both in internal dimensions and in the geometrical data of the dual CFT. As a concrete example, we examine in section 4 how the optical depth depends on the size, for the standard case of a thermal sate in maximally supersymmetric Yang--Mills theory on a toroidal box ${\bf T}^d$. Finally, we point out that the T-duality operating in this example allows us to `derive' the matrix-model description of a single thermal cell, as well as interesting consequences for the behavior of D-brane probes.

\newsec{Optical Depth Of Holographic Thermal Atmospheres}

\noindent

In this section we recall the estimate done in \refs\uschaos\ relating the free-fall time scale across the near-horizon region with the notion of `optical depth' in backgrounds with holographic interpretation. 

We may  simplify the discussion, with no crucial loss of generality,  by considering a geometry composed of two regions: we have an asymptotic AdS$_{d+2}$ geometry of curvature radius $b$,  joined at an energy scale $T \sim r_0 /b^2$ to a Rindler geometry, which is subsequently cut-off by a Planckian stretched horizon. This system describes a large-$N$ CFT  on the non-dynamical $(d+1)$-dimensional static spacetime 
\eqn\qftm{
ds^2_{\rm CFT} = -dt^2 + d\ell_d^2 \;,
}
on a thermal state at temperature $T$. We allow for finite-$N$ effects by a phenomenological procedure, i.e. by stretching the horizon to a Planckian layer of strong coupling, thereby introducing a finite number of degrees of freedom of order $N_{\rm eff} \sim (b/\ell_{\rm P})^d$. We thus have a metric of the form
\eqn\asym{
ds^2 \approx {r^2 \over b^2} \left(-dt^2 + d\ell_d^2 \;\right) + {b^2 dr^2 \over r^2}\;, \qquad r\gg r_0
\;,
}
joined at $r\sim r_0 \sim b^2 T$ to a Rindler patch
\eqn\rindlp{
ds^2 \approx  -(2\pi T)^2 \rho^2\,dt^2 + d\rho^2 + (Tb)^2 \,d\ell_d^2 \;, \qquad \rho\ll \rho_\beta \;.
}
The two portions of the metric are matched smoothly at $r_\beta = r_0 + {b^2 \over 4\pi \beta}$, or $\rho_\beta = b/2\pi$. A stretched horizon sits at Planck distance from the horizon, $\rho_* \sim \ell_{\rm P}$. 

The optical manifold with metric $ds^2_{\rm op} \equiv ds^2 /|g_{00}|$ has two characteristic pieces as well. The asymptotic region $r\gg r_0$  is mapped to a flat strip of thickness $\beta$,
\eqn\opas{
ds^2_{\rm op} \approx -dt^2 + dz^2 + d\ell_d^2 \;, \qquad 0<z<z_\beta \sim \beta\;, }
in terms of the Regge--Wheeler coordinate, $z=b^2/r$. On the other hand, the near-horizon region $r\sim r_0$ is locally given by a static hyperboloid,
\eqn\neaop{
ds^2_{\rm op} \approx -dt^2 + dz^2 +  e^{4\pi T (z-z_\beta)} d\ell_d^2\;, \qquad z>z_\beta \;,}
with radial optical thickness
\eqn\othi{
z_* - z_\beta = {1\over 2\pi T} \log\left({\rho_\beta \over \rho_*}\right) \sim \beta\,\log\left({b \over \ell_{\rm P}}\right)\;.
}
Since the optical depth determines the free-fall time $\tau_*$ of a photon across the Rindler region, we have, up to $\CO(1)$ numerical factors
\eqn\finf{
\tau_* \sim \beta\,\log (N_{\rm eff})\;,
}
since $N_{\rm eff} \sim (b/\ell_{\rm P})^d$ is the number of degrees of freedom of the dual CFT. For non-conformal theories it gives a sort of `running' central charge with non-trivial scale dependence. A good example is provided by the
gravity duals of the D$p$-brane theories with $p<5$, where 
$$
N_{\rm eff} (T) = N^2\,(\lambda_p \,T^{\,p-3} \,)^{p-3 \over 5-p}\;,
$$
with $\lambda_p$ the dimension-full 't Hooft coupling of the $(p+1)$-dimensional Yang--Mills theory \refs\itzhaki. It is remarkable that, despite the non-AdS nature of the asymptotic regions, the optical metric is still given by
\opas\ for all D$p$-brane throats with $p<5$, including the fact that the optical distance to infinity is given by the
inverse temperature $\beta$. 

The optical depth is readily interpreted as a free-fall time to the stretched horizon. The same time scale
is associated to the Ohm diffusion at the stretched horizon, induced by moving charges in the near-horizon region (cf.   \refs{\Damour, \Thorne, \Susskindbook}). This is particularly clear in the optical representation of the thermal cell depicted in Figure 1. The whole non-compact asymptotic region is mapped to a small box of size $\beta$, so that induced charges at the stretched horizon are quite insensitive to the motion of source charges in the asymptotic region. The optical metric is dominated by the near-horizon region, and the time-scale for global causal communication across this region defines the time scale for large-scale rearrangements of induced charges at the stretched horizon (notice that Maxwell's equations are conformally invariant, so that electromagnetic field solutions can be faithfully studied in the optical frame).

 \bigskip
\centerline{\epsfxsize=0.5\hsize\epsfbox{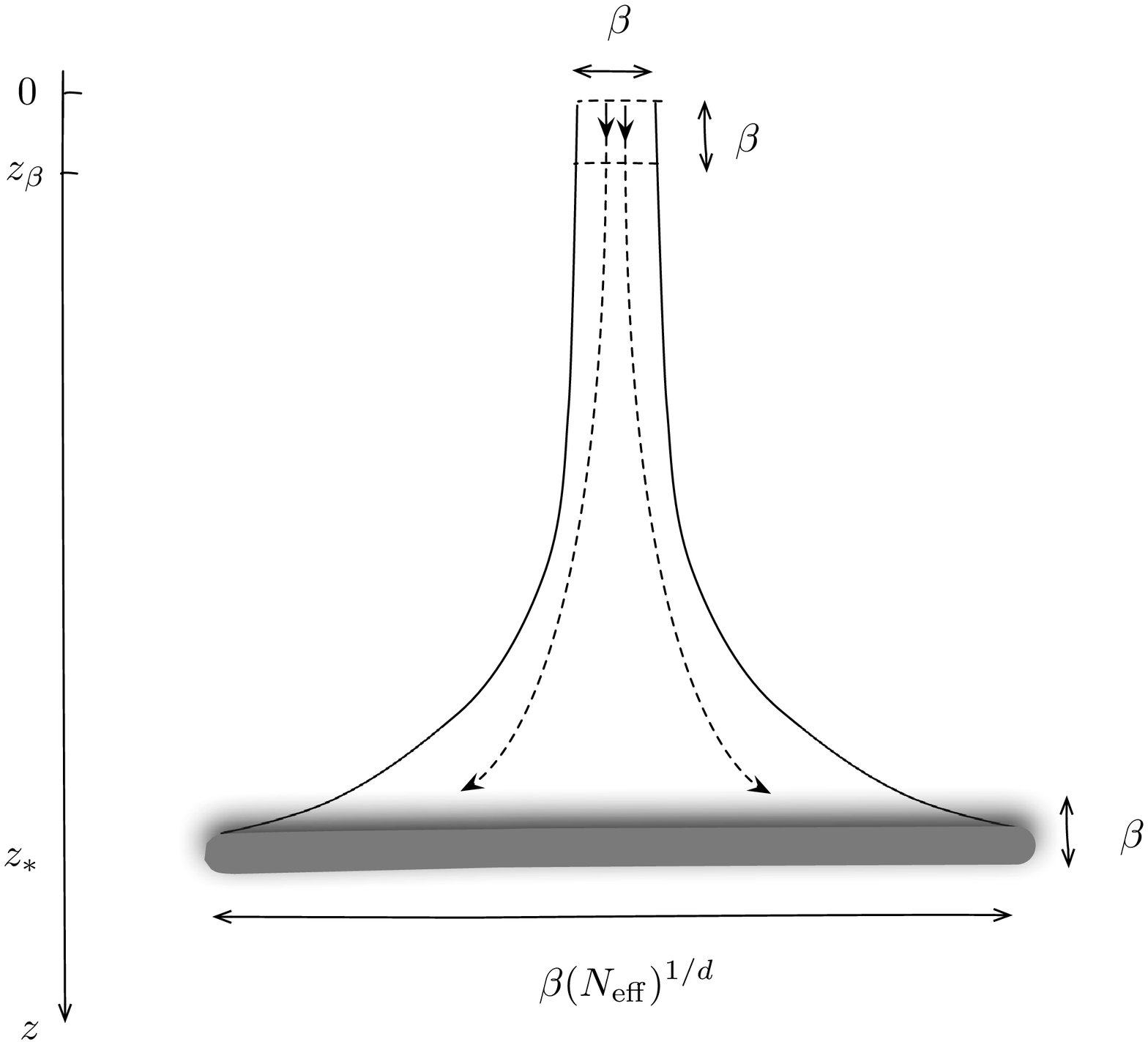}}
\noindent{\ninepoint\sl \baselineskip=8pt {\bf Figure 1:} {\ninerm
The optical box of a single thermal cell of CFT volume $\beta^d$, drawn to indicate the optical-volume expansion of the spatial sections up to an optical volume $ N_{\rm eff}$ times larger at the stretched horizon, itself with optical thickness of $\CO(\beta)$. The vacuum piece $z\ll z_\beta$ has negligible optical volume compared to the Rindler piece $z_\beta \ll z\ll z_*$, which dominates the AdS/CFT computation of any boundary observable in the long-time limit. Exponential sensitivity to initial conditions in the UV leads to chaotic classical behavior in the near-horizon region. }}
\bigskip

\newsec{Hidden Structure And Small Scramblers}

\noindent

Our computation of the optical depth in \finf\  assumes that \asym\ is presented in the Einstein frame and in the minimal holographic form, i.e. we have a $(d+2)$-dimensional bulk to represent a $(d+1)$-dimensional CFT at large $N_{\rm eff}$. Refinements of the AdS/CFT correspondence often involve the discussion of compact factors in the bulk. 
In this case, the question arises as to what definition of the Planck length must be used in the two places where it appears, namely in the location of the stretched horizon and in the normalization of the entropy. We may use either the $(d+2)$-dimensional definition, or the $(d+k+2)$-dimensional definition in a background with a compact $k$-dimensional factor. A natural criterion would demand that we `integrate out' the extra compact factors when their local size at the near-horizon region is smaller than the local red-shifted inverse-temperature. On the other hand, at temperatures large enough to `see' the extra compact cycles we should use the higher-dimensional picture (see \refs{\exten, \kogan, \ncop} for an `extensive' use  of this criterion in  related contexts.) 

A large class of holographic backgrounds can be parametrized as `warped products' of AdS with a compact manifold of the same overall (but positive) curvature. 
To be more precise, consider Einstein-frame metrics parametrized in the form
\eqn\warpi{
ds^2 = {r^2 \over b(r)^2 } \left(-h(r)dt^2 + d\ell_d^2 \,\right) + {b(r)^2 \over r^2} {dr^2 \over h(r)} + b(r)^2 \,dy_k^2\;,
}
where $b(r)$ is a function characterizing the overall curvature, or order $1/b(r)^2$ at radius $r$. The  coordinates $y_k$ parametrize a $k$-dimensional compact factor ${\bf K}_k$ of $\CO(1)$ curvature, warped by the profile function $b(r)$.  The function $h(r)$ is the thermal factor, admitting a near-horizon parametrization  $h(r) \sim (r-r_0)/r_0$ in order of magnitude, so that the Hawking temperature reads $T\sim r_0 /b_0^2$, with $b_0 \equiv b(r_0)$.  More generally, we have an approximate UV/IR relation for the energies measured with respect to the $t$ variable,  $E(r) \sim r/b(r)^2$, as corresponds to the
warped product  of the local approximate form $\left[{\rm AdS}_{d+2}\right]_r \times \left[{\bf K}_k\right]_r$.  Interestingly, most AdS/CFT backgrounds whose geometry is determined by a generic deformation of an UV fixed point admit a representation of the form \warpi, since a single dominant relevant operator will determine a geometry with a single overall curvature scale, both in compact and non-compact factors. 

Repeating the calculation of the near-horizon optical depth for \warpi\ we find \foot{The optical distance from the edge of the thermal atmosphere to the boundary at $r=\infty$ is of $\CO(\beta)$ for any background of the form 
\warpi,  provided $b(r)$ has power-like dependence on $r$. A notable exception to this rule is the case of the NS5-brane throat, dual to the Little String Theories (LST), for which the free fall time from infinity is of order $\beta_{\rm H} \,\log(S_{\rm max})$, with $\beta_{\rm H}$ the inverse Hagedorn temperature of the LST and $S_{\rm max}$ a cutoff value of the entropy (cf. \refs\lstus.) }
$
\tau_* = z_* - z_\beta \sim \beta\,\log\left({b_0 /{\bar  \ell}_{\rm P}}\right)
$, 
where ${\bar \ell}_{\rm P}$ denotes the Planck length in $d+k+2$ dimensions. 
On the other hand, the entropy in a single thermal cell reads
\eqn\enstc{
\scell ={S \over \ncell} \sim {1\over VT^d} {1\over {\bar \ell}_{\rm P}^{\;d+k}} \left({r_0 \over b_0}\right)^d V \cdot V_k (r_0)\;.
}
Using $V_k (r_0) \sim b_0^k$ and $T\sim r_0 /b_0^2$ we find 
$
S_{\rm cell} \sim  
\left({b_0 / {\bar \ell}_{\rm P}}\right)^{d+k}
$, 
so that the main result 
\eqn\mrse{
\tau_* \sim \beta \,\log\,(\scell)
}
 is obtained in full generality, up to $\CO(1)$ coefficients. Our crucial observation here is that \mrse\ is obtained  independently of whether we use the $(d+k+2)$-picture or the $(d+2)$-dimensional picture with effective Planck length ${ \ell}_{\rm P}$, obtained by Kaluza--Klein reduction  of the compact factor at the horizon ${\bf K}_k (r_0)$. The two definitions of  Planck length satisfy  ${ \ell}_{\rm P}^{\;d} = ({\bar \ell}_{\rm P})^{d+k} / b_0^k$, which in turn implies
 $
 \left({b_0 /{\bar \ell}_{\rm P}}\right)^{d+k} = \left({b_0 / { \ell}_{\rm P}}\right)^d
 $, 
 thus ensuring \mrse\ also in the $(d+2)$-dimensional Einstein frame. 

\subsec{Zooming Inside The Thermal Cell}

\noindent

A most interesting fact about this result is the generalization of \mrse\ to the degenerate case of $d=0$, i.e. the situation where the
dual system is purely quantum mechanical and the horizon is only extended in `internal' compact dimensions, so that the previous notion of `thermal cell' is not defined.  In this case one has no alternative to using the $(d+k+2)$-dimensional description, whose optical depth scales with the logarithm of the {\it total} entropy $S\sim (b_0 / {\bar \ell}_{\rm P})^k$: 
\eqn\rnod{
\tau_* \sim \beta\,\log\,(S)\;.
} 
A characteristic example of this behavior is  the 
 four-dimensional Reissner--Nordstrom black hole, with near-horizon geometry AdS$_{1+1} \times {\bf S}^2$. In this case  the optical depth scales with the logarithm of the full entropy,
despite the fact that the size $b_0$ of the ${\bf S}^2$ is much smaller than the inverse temperature $\beta$ in the extremal $T\rightarrow \infty$ limit. Notice that this result for the optical depth must be distinguished from that of the Schwarzschild black hole, despite involving the same formula. The reason is that Schwarzschild black holes have a size of the same order as the thermal length, for any temperature, so that both \rnod\ and \mrse\ apply to them.

This example suggests that the law \mrse\ should be replaced by \rnod\  when the system is smaller than the inverse temperature scale. It is interesting to consider examples in which the reduction to a single thermal cell can be achieved by varying a continuous  control parameter. The simplest possibility is that of a CFT on a  sphere  ${\bf S}^d $ of radius $R$. The bulk representation in the high-temperature phase is a large AdS black hole. For $T\sim 1/R$ the black hole has size of $\CO(1)$ in units of the AdS curvature and we have a CFT living on a single thermal cell. For $T\ll 1/R$, which corresponds to a system `smaller' than a single thermal cell, the dominant background is the vacuum AdS manifold, with no horizons. The propagation of bulk signals inside global AdS occurs on the time scale of the order of the curvature radius, i.e. we have a purely `ballistic' regime for signal propagation on the CFT sphere (recall that the interactions with bulk radiation are down by
  one power of $1/N^2$.) This is all natural since the finite size of the sphere gaps the spectrum of the CFT and we only see the vacuum on scales smaller than the thermal length. We conclude that the Hawking--Page transition makes the transition between \mrse\ and \rnod\ somewhat degenerate in this case, since all low-$T$ entropies are of $\CO(1)$ in the large-$N$ expansion.  

A more interesting situation would apply if the theory conserves $\CO(N_{\rm eff})$ worth of degrees of freedom when system is smaller than a thermal cell. In this case, we need to realize an entropy of $\CO(N_{\rm eff}) \sim \CO(N^2)$ in a purely quantum mechanical system, i.e. as a black-hole metric of the form \warpi, with $d=0$. This physical constraint, combined with the Hawking--Page transition in the gapped case, suggests that we consider large-$N$ phase transitions with the bulk interpretation of topological jumps between metrics of the form \warpi\  with a different partition of `internal' and `spacetime' directions.

To be more specific,
consider the induced geometry at fixed radial variable $r$ and fixed time $t$, with topology ${\bf V}_d \times {\bf K}_k$, where ${\bf V}_d$ denotes the spatial section of the QFT metric \qftm. A topological flop into a system with $d=0$ has the schematic form
${\bf V}_d \times {\bf K}_k \rightarrow {\bf K}_{d+k}$, and will be likely to occur when the three manifolds have about the same proper size at the radius scale set by the horizon $r\sim r_0$, corresponding to a temperature $T\sim r_0 / b_0^2$. The size of ${\bf K}_k (r_0)$ is given by $b_0$, whereas the size of ${\bf V}_d (r_0) $ is or order 
 $L \cdot r_0 / b_0 \sim (LT) b_0$, with $L$ the size of ${\bf V}_d$ in the QFT metric. Using the UV/IR relation, these sizes are about equal for $LT \sim 1$, i.e. when the system contains a single thermal cell. 
 
 We expect the quantum-mechanical phase to dominate in the low-temperature regime, $LT\ll 1$, when the system is smaller than a single thermal cell. This is natural since the entropy and/or free energy of the QFT is computed in the bulk prescription by evaluating volume integrals as a function of  $LT$ for the two manifolds. At the transition one has $S\sim S_{\rm cell}$, with \mrse\ holding at $LT\gg 1$ and \rnod\ taking over for $LT\ll 1$.

\newsec{D-branes And The Matrix Model Of A Single Thermal Cell}

\noindent

A particular example of the conjectured topological flops described in the previous section can be studied in great detail by using
the the very explicit solutions of D-brane backgrounds in type II string theories. 
According to basic AdS/CFT lore, strongly coupled SYM theories in $d+1$ dimensions admit a bulk gravity dual description based on the near-horizon (string frame) metric/dilaton of D$d$-branes \refs\itzhaki, 
\eqn\dppp{ ds^2_{{\rm D}d} = {1\over \sqrt{H_d}} \left(-h(r) dt^2 + d\ell_d^2 \right) +  \sqrt{H_d} \left( {dr^2 \over h(r)} + r^2 d\Omega_{8-d}^2 \right) \;, \qquad e^{-2(\phi-\phi_\infty)} =( H_d)^{d-3 \over 2} \;,}
where $H_d = (R_d / r)^{7-d}$ and $h(r) = 1- (r_0 /r)^{7-d}$. This background is a particular case of \asym, up to a change of coordinates, a Kaluza--Klein reduction on the compact ${\bf S}^{8-d}$ sphere,  and a final rescaling to an Einstein frame metric.\foot{Notice that the D-brane background is of the form \warpi, both in string frame and in Einstein frame.} For $\ell_d \in {\bf T}^d$, a torus of   size $L$, the associated winding modes become light as $r\rightarrow 0$. Resolving this singularity via a T-duality on the ${\bf T}^d$  one finds a metric which becomes unstable to  localization (a global version of \refs\gl)  for $r\ll \alpha' /L$ (see for example \refs\kogan\ for a detailed account). The localized metric is then that of D0-branes, i.e. \dppp\ with $d=0$. 

The T-duality transition occurs at the point where local winding modes become of stringy mass, equivalently the
local proper size of the spatial torus in the string-frame metric \dppp\  is of order
$$
L(r_{\alpha'}) \sim L\left({R_d \over r_{\alpha'}}\right)^{7-d \over 4} \sim \sqrt{\alpha'}\;.
$$
The T-dual geometry for $r<r_{\alpha'}$ contains a  torus growing towards small $r$,   supporting a uniform distribution of D0-branes. This geometry is globally unstable through the 
 topological `flop'   ${\bf T}^d \times {\bf S}^{8-d} \rightarrow {\bf S}^8$. The local size of the warped ${\bf S}^{8-d}$ is proportional to $r$, while that of the T-dual torus is proportional to ${\tilde L} = \alpha' /L$. This determines the location of the flop at $r_{\rm flop} \sim {\tilde L}$, where both fibers have roughly the same size,  so that they can have the same action as a round ${\bf S}^8$.  Since $N_{\rm eff}$ is proportional to the entropy, which in turn scales with the volume of the fibers, our construction guarantees that \mrse\ continues to apply in order of magnitude  across this transition, where $S_{\rm cell}$ is interpreted in the low-temperature regime as the entropy of the large-$N$ quantum mechanics of the D0-brane system.\foot{The entropy computed through the Bekenstein--Hawking formula is invariant under T-duality and also under S-duality.}  So we have a system where we go from a law of type \mrse\ to a law of type \rnod, with the difference that 
 the low-temperature entropy still shows non-trivial $T$-dependence.

 \bigskip
\centerline{\epsfxsize=0.4\hsize\epsfbox{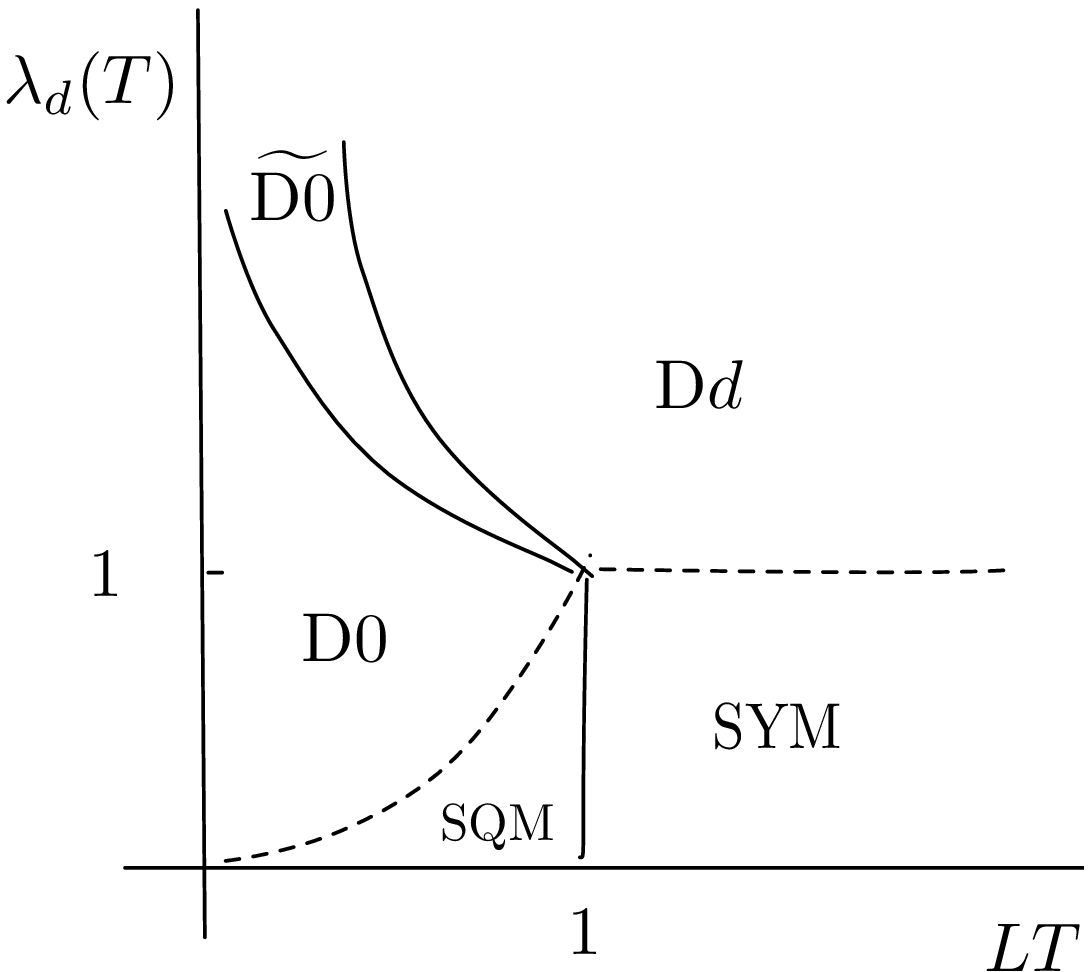}}
\noindent{\ninepoint\sl \baselineskip=8pt {\bf Figure 2:} {\ninerm 
Phases of the D$d$-brane system as a function of $LT$ and the effective dimensionless coupling in $d+1$ dimensions. In the gravity regimes above the dotted line, we go from the D$d$-brane geometry to the smeared D0-brane geometry, $\widetilde{{\rm D}0}$, as the torus size goes through the T-duality transition. For even smaller sizes we have a localization transition to the metric of localized D0-branes. The large-$N$  thermodynamic functions   are T-duality invariant, and  undergo a first-first order phase transition without $\CO(N^2)$ latent heat at the localization curve. }}
\bigskip

It is instructive to  express these results in terms of SYM variables. Let $\lambda_d = g_{\rm YM}^2 N$ be the 't Hooft coupling of the SYM theory in $d+1$ dimensions. It has length dimension $d-3$ and a dimensionless coupling characterizing the intrinsic strength of interactions at the energy scale $T$ is the running coupling $\lambda_d (T) = \lambda_d \,T^{d-3}$.

The SYM theory can be described by the holographic model \dppp\ provided it is sufficiently strongly coupled, i.e. 
for $\lambda_d (T) \gg 1$, and the temperature is large enough. At temperatures below the critical line $\lambda_d (T) \sim (LT)^{2(d-5)}$, corresponding to $r_0 \sim r_{\alpha'}$,  the metric \dppp\ must be substituted by the T-dual background of $N$ smeared D0-branes. Further down in temperature the horizon reaches the critical stability line $r_0 \sim r_{\rm flop}$, corresponding to $\lambda_d (T) \sim (LT)^{d-5}$ in YM variables, where  the system localizes to the large-$N$ quantum mechanics of $N$ coincident  D0-branes. In order to better represent this behavior, it is useful to define an effective renormalized thermal length $\ell_T$, by the relation
\eqn\ntle{
\ell_T^{\,5-d}= {\beta^{\,5-d} \over \lambda_d (T)} \;,}
where strong coupling effects $\lambda_d (T) \gg 1$ make it smaller than the perturbative notion of thermal length of $\CO(\beta)$. 
The expression
   $$\tau_* \sim \beta \log(N_{\rm eff})$$ is then valid  at all temperatures satisfying the strong coupling condition $\lambda_d (T) \gg 1$, where  the effective number of degrees of freedom runs as that of strongly-coupled $(d+1)$-dimensional SYM:
$$
N_{\rm eff} (T) = N^2 \left(\lambda_d (T)  \right)^{d-3 \over 5-d} \sim (S_{|{\rm D}d})_{\rm cell}
$$ when the size of the system is larger than the effective thermal length
$L\gg \ell_T$. For tori smaller than the effective thermal length, $L \ll \ell_T$, we cross-over to  the effective number of degrees of freedom of the large-$N$ quantum mechanics  at strong coupling: 
$$
N_{\rm eff} (T) =N^2  \left(\lambda_0 (T)  \right)^{-{3\over 5}} \sim S_{| {\rm D}0}\;,
$$
with $\lambda_0 (T) = \lambda_0 \,T^{-3} $ the effective dimensionless coupling of the D0-brane system. Using 
the Kaluza--Klein reduction formula for the SYM theory on the torus, we can write $\lambda_0 = \lambda_d /L^d$, which allows us to express the effective D0 coupling in terms of the original effective Yang--Mills coupling in $d+1$ dimensions through the relation $\lambda_0 (T) = \lambda_d (T)(LT)^{-d}$. Notice that the two expressions for $N_{\rm eff}$ match in order of magnitude across the critical transition line $L\sim \ell_T$. 
Hence, the behavior of a near-extremal black hole is obtained, up to an effective shrinkage of the thermal length by effects attributable to the large value of the 't Hooft coupling.

 \bigskip
\centerline{\epsfxsize=0.4\hsize\epsfbox{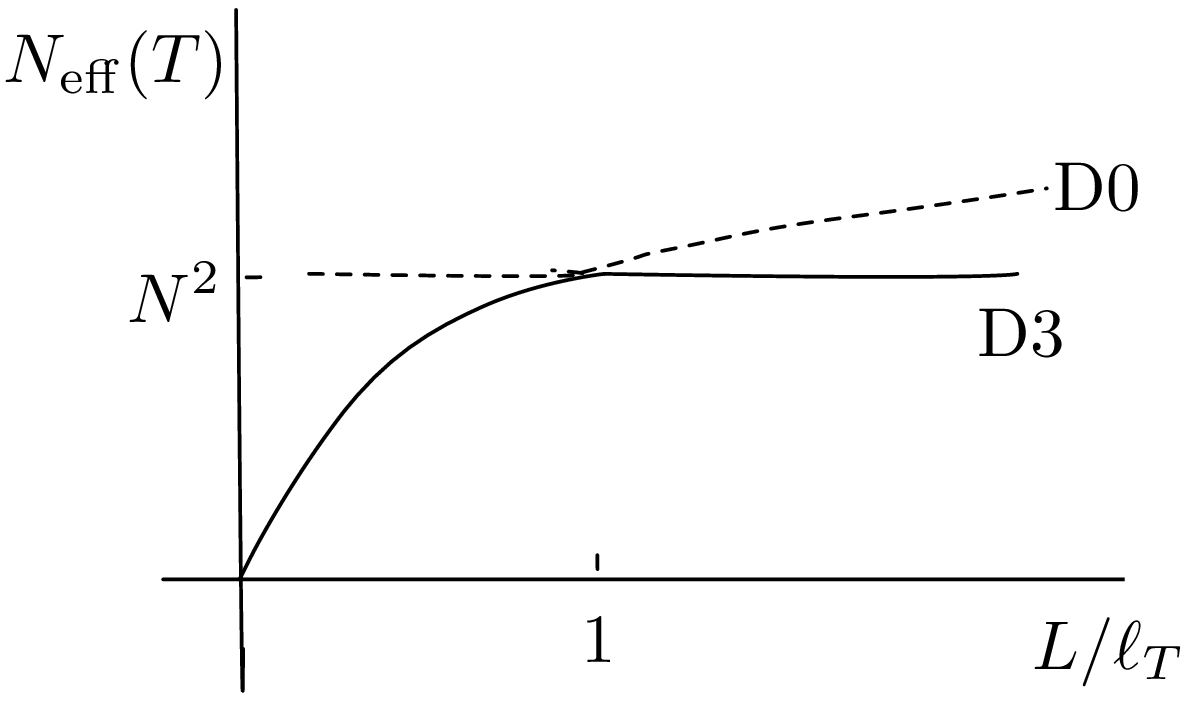}}
\noindent{\ninepoint\sl \baselineskip=8pt {\bf Figure 3:} {\ninerm
Plot of the temperature-dependent value of $N_{\rm eff}$ (full line) as a function of the system size for the SYM$_{3+1}$ theory on a torus. The transition from $N_{\rm eff} = S_{\rm cell}$ to $N_{\rm eff} = S$ occurs at the localization curve, defined by $L\sim \ell_T$. Notice that the dominant solution {\it minimizes} $N_{\rm eff}$, even if it maximizes the total entropy.
 }}
\bigskip

These results can be interpreted by saying that, as far as scrambling properties is concerned, we can regard
the large-$N$ gauge theory as an effective lattice matrix model, with one lattice site per thermal cell (as first suggested in \refs\Sekino). The scrambling proceeds diffusively along the lattice, but the $\CO(N_{\rm eff})$ degrees of freedom within each site are scrambled as in a single-site matrix model. In the concrete example at hand, the matrix model in question is determined  completely by T-duality  to be that of \refs\Matrix, i.e. the D0-branes matrix system.  Notice that, once
we establish this scenario for the geometrical notion of optical length, the chaotic kinetic model of \refs\uschaos\ automatically guarantees the same for the scrambling time. 

\subsec{T-duality And Probe Locality}

\noindent

The result \mrse\ stands out for its  universal character. It is interesting to note that the free-fall time of other probes, such as near-horizon extended branes, follow the same pattern.  
Let us consider a CFT-space filling brane falling rigidly in the bulk  metric. The action depends on  an effective tension $\sigma$ and an charge $q$ through  
\eqn\tchar{
I_{\rm brane} = I_{\rm NG} + I_{\rm WZ} = -\sigma {\rm Vol}\left[\Sigma\right]  + q  \int_{\Sigma}  A_{\rm WZ} \;,
}
where $\Sigma$ is the word-volume  of the brane and $A_{\rm WZ}$ is a Wess--Zumino field coupled minimally to the brane and assumed to be smooth at the horizon. Transforming to Regge--Wheeler coordinates and approximating the action in the Rindler region one finds 
\eqn\barw{
I_{\rm brane} \approx  -\int dt \left(m_{\rm eff} (z) \sqrt{1-{\dot z}^2} + v_0\right)\;,
}
where $v_0$ is a constant and 
$$
m_{\rm eff} (z) \approx m_0 \,e^{-2\pi T (z-z_\beta)}\;,
$$
for some positive constant $m_0$. We see that the effective mass of the brane vanishes exponentially as we approach the horizon. Hence, the fall time from $z=z_\beta$ to the stretched horizon is of order  $z_* - z_\beta = \tau_*$. 

While the free-fall time of branes across the near-horizon region is given by the optical depth, it is less obvious that this time scale is directly related to a scrambling process in this case, since the probe is already `delocalized' and not prone to kinetic-type arguments. It is rather more natural to view the scrambling of such probes in the context of matrix model scrambling, along the proposal of \refs\Sekino. 

The use of T-duality in the discussion of the previous section introduces an interesting angle on the question of probe locality. We can expect, on general grounds, that scrambling will be the hardest to accomplish for very localized probes, in which case the chaotic dynamics suggested in \refs\uschaos\ is summoned to perform the feat. 
A paradigm of a localized probe is certainly a low-dimensional D-brane. 

For example, starting with five-dimensional Yang--Mills theory, realized on a stack of D4-branes in type-IIA string theory, we may consider  a probe D0-brane falling in the near-horizon region and scattering from the stretched horizon. If we roll the D4-branes on a 4-torus and take the limit of a single thermal cell, the T-dual picture involves a localized state of D0-branes which have undergone  localization from a stack of smeared D0-branes, themselves T-dual of the original wrapped D4-branes. 

Under the T-duality, the probe D0-brane becomes a probe D4-brane. Hence, after the localization we have a thermal state of D0-branes being probed by an external D4-brane whose wave-function has no structure as a function of a local coordinate on the horizon of the D0-branes.  Conversely, had we started with a D4-brane probe falling into the original black D4-brane system, the T-duality would take us to a D0-brane probe falling into the near-horizon region of the D0-brane black hole. 

While a localized probe of D0-brane type can naturally scramble via the kinetic model of \refs\uschaos, it is 
less obvious how to apply these ideas to pure matrix-model scrambling, when the probe is an extended brane of the same dimension as the black-branes of interest. However, the action of T-duality suggest that the same scrambling time scale should apply to both situations, lending credence to the expectation that one single scrambling time scale, $\tau_*$, governs all types fast scrambling at horizons.

 \newsec{Conclusions}
 
 \noindent
 
 We have explored aspects of the concept of `optical depth' introduced in \uschaos\ as a measure of fast scrambling in generic horizons. We have studied its behavior under large-$N$ phase transitions triggered by topological jumps in bulk AdS/CFT backgrounds, and associated to finite-size effects in the CFT.
In particular, a very precise picture emerges for the case of SYM theory on ${\bf T}^d$, a $d$-dimensional torus of size $L$. As the torus  shrinks down to the scale of a (renormalized) thermal cell, a localization phase transition takes place across the parameter line defined by the equality of the total and the cell entropies, i.e. $S\sim S_{\rm cell}$. At the localization transition point, the original system is described holographically in terms of the T-dual of the original brane state, including possible D-brane probes used to monitor the scrambling.

These arguments show that large $N$ thermal states of strongly-coupled CFTs may be regarded, as far as scrambling properties is concerned, as lattices of matrix models with rank $N_{\rm eff} (T) \sim S_{\rm cell}$ and a lattice  spacing of one thermal length. The model based on D-branes determines explicitly the single-cell effective matrix model by means of the topological phase transition mentioned above. 

In this picture, diffusive scrambling with time step $\tau_*$ would operate over length scales in excess of one thermal length, leading to a scrambling time $\tau_s \sim \tau_* (n_{\rm cell})^{2/d}$, whereas genuine fast scrambling would be associated to the single-cell level, and interpreted in the CFT as an anomalous `large' scrambling time $\tau_* \sim \beta \,\log(N_{\rm eff})$.  Hence, we can say that, according to these results, fast scramblers are necessarily {\it small}, in the appropriate sense. 

Another further consequence of the D-brane model is the suggested T-duality between the scrambling of a localized probe, by the chaotic kinetic process of \refs\uschaos, and the more mysterious matrix-model scrambling expected at the level of color degrees of freedom, along lines of \refs\Sekino. More generally, it should be interesting to study the fast thermalization properties of matrix models along the lines of \refs\festuccia\ and more specifically the recent work \refs\beren.

\bigskip{\bf Acknowledgements:}  We are indebted to P. Hayden and E. Rabinovici for useful discussions, as well as the participants and organizers of the Aspen Workshop ``Quantum Information in Quantum Gravity and Condensed-Matter physics". J.L.F.Barb\'on wishes to thank the Aspen Center for Physics and the University of Oviedo for hospitality while this work was completed.
 This work  was partially supported by MEC and FEDER under a grant FPA2009-07908, the Spanish
Consolider-Ingenio 2010 Programme CPAN (CSD2007-00042) and  Comunidad Aut\'onoma de Madrid under grant HEPHACOS S2009/ESP-1473. J.M.M. is supported by a FPU fellowship from MICINN.

{\ninerm{
\listrefs
}}

\bye